\documentclass[11pt,twoside]{article}

\usepackage{asp2006}
\usepackage{epsf}
\usepackage{psfig}
\usepackage{lscape}
\usepackage{natbib}
\usepackage{astrojournals}

\begin{document}

\newcommand{\um}{\,$\mu$m}

\title{Laboratory studies of polycyclic aromatic hydrocarbons: the search for
interstellar candidates}   
\author{C. Joblin$^{1,2}$, O. Bern\'e$^{1,2}$, A. Simon$^{1,2}$, G. Mulas$^{3}$}   
\affil{1- Universit\'e de Toulouse ; UPS ; CESR ; 9 av. colonel Roche, F-31028 Toulouse cedex 9, France\\
2- CNRS ; UMR5187 ; F-31028 Toulouse, France\\
3- INAF~\textendash~Osservatorio Astronomico di Cagliari~\textendash~Astrochemistry  Group,
Strada 54, Loc. Poggio dei Pini, I\textendash09012 Capoterra (CA), Italy}    

\begin{abstract} 
Polycyclic Aromatic Hydrocarbons (PAHs) are considered as a major constituent
of interstellar dust. They have been proposed as the carriers of the Aromatic Infrared
Bands (AIBs) observed in emission in the mid-IR. They likely have a
significant contribution to various features of the extinction curve such as the 220 nm bump,
the far-UV rise and the diffuse interstellar bands.
Emission bands are also expected in the far-IR, which are better
fingerprints of molecular identity than the AIBs. They will be searched for with the
Herschel Space Observatory. Rotational emission is also expected in the mm range
for those molecules which carry significant dipole moments. 

Despite spectroscopic studies in the laboratory, no individual PAH species
could be identified. This emphasises the need for an investigation on where interstellar
PAHs come from and how they evolve due to environmental conditions: ionisation
and dissociation upon UV irradiation, interactions with electrons, gas and dust.
Insights into this question have recently been obtained  from the analysis of the AIB
spectra in different environments including mild UV-excited photodissociation
regions, planetary nebulae and protoplanetary disks. PAH species are found -(i)- to be
produced by destruction of very small grains (VSGs), -(ii)- to have an evolving charge
state: neutral or cationic, and -(iii)- to undergo severe destruction in highly UV-irradiated
environments, only the largest species being able to survive.
There is also evidence for PAH species to contribute to the depletion of heavy atoms
from the gas phase, in particular Si and Fe.

This paper illustrates how laboratory work can be inspired from observations.
In particular there is a need for understanding the chemical properties of PAHs and
PAH-related species, including VSGs, in physical conditions that mimic
those found in interstellar space.
This motivates a joint effort between astrophysicists, physicists and chemists.
Such interdisciplinary studies are currently performed, taking advantage of the
PIRENEA set-up, a cold ion trap dedicated to astrochemistry.

\end{abstract}

\section{Introduction} 
Polycyclic Aromatic Hydrocarbons (PAHs) are usually considered as the carriers
of the Aromatic Infrared Bands (AIBs) observed in emission between 3 and 13\um\
in many astronomical objects that are submitted to UV photon
irradiation: the so-called photodissociation
regions \citep[PDRs;] [] {leg84, all85}.
They are expected to contribute significantly to various features
of the UV-visible extinction curve such as the 220 nm bump, the far-UV rise
\citep{job92b} and the diffuse interstellar bands \citep{sal99}.
Emission bands are also expected to occur in the far-IR, that are
better fingerprints of molecular identity than the mid-IR bands \citep{job02a, mul06c}.
They will be searched for with the Herschel Space Observatory (HSO).

PAHs (the carriers of the AIBs) appear as a major component in
the cycle of dust in the Milky Way from evolved stars (planetary nebulae),
to the diffuse interstellar medium (ISM) and bright PDRs associated to molecular
clouds and the circumstellar environment of young stars including their disks.
Their emission is also well observed in external galaxies
\citep[e.g.][]{smi07, spo07, gal08a, gal08b}
and even in ultra-luminous IR galaxies (ULIRGs) at z=1-3 \citep[e.g.][]{yan05, saj07}.
AIB spectra show characteristic features which fall at 3.3, 6.2, "7.7", 8.6, 11.3 and 12.7\um.
The observed spectra exhibit variations in band positions, profiles and relative
intensities and also differ in their satellite features.
Two general approaches can, therefore, be adopted to use these spectral features
as diagnostics of the chemical and physical conditions from galaxies to circumstellar
environments.
One consists in searching for groups of astronomical objects based on classification of
the spectral features \citep{pee02,van04}.
The other one is based on a better understanding of the nature of the AIB carriers
and their physical and chemical evolution due to environmental conditions.
This approach is guided by experimental and theoretical studies
\citep[cf. for instance][] {lan96, pau01,hud05, mul06a, bau08, joa09}.
For instance, it has been shown that the 7.7/11.3\um\ band intensity ratio
can be used as a tracer of the PAH$^+$/PAH$^0$ ratio. The value of this
ratio depends on the competition  between ionisation and recombination
with electrons,  which involve the UV flux density and  the electron density,
respectively \citep[cf. Figure~6.7 in] []{tie05}.
This band ratio was used in a large sample of galaxies to characterise
the local physical conditions \citep{gal08b}.

The search for the identification of individual PAH molecules or families of species
in the AIB spectra, has not been successful so far.
This lack emphasises the need for an investigation on where
interstellar PAHs come from and how they evolve due to environmental
conditions: ionisation and dissociation upon UV irradiation, destruction in
shocks, interactions with electrons, gas and dust. These questions motivate
fundamental studies on analogues of interstellar PAHs in physical conditions
that mimic those found in the ISM. This leads to the
development of specific laboratory set-ups, such as PIRENEA, a cold
ion trap dedicated to astrochemistry.

\section{What can we learn from mid-IR observations?\\
A chemical evolution scheme for the mid-IR emission carriers }
\label{IRobs}

Since the IRAS mission, it is well known that there exists a population of very small particles which
are transiently heated by the absorption of single UV photons and emit in the mid-IR
range. PAHs were proposed to account for the AIBs and another component, called very small
grains  \citep[VSGs;] [] {des90}, was introduced as the carrier of the diffuse IR
emission in the 25\,$\mu$m IRAS band.
In the last years, space missions, and in particular the European
Infrared Space Observatory and the NASA Spitzer Space Telescope, have gathered
a wealth of spectral data that allows us to better characterise the various mid-IR emission
features and their variations. One can consider either a large sample of
individual objects that are not spatially resolved, or study the spatial variations 
of these bands within extended objects.
In the first approach, the objects can be quite compact which implies that the 
spectrum integrated in the spectrometer beam is likely to include 
the contribution from different types of regions. The exciting 
central source may also be difficult to characterise (e.g. PNe, compact
\ion{H}{ii} regions, disks, etc.). In the case of
extended objects on the other hand, one can hope to access chemical frontiers
(gradients in the abundances of species).

Spectro-imagery  of extended PDRs has become possible in the mid-IR (5-16\,$\mu$m)
with ISOCAM in CVF mode \citep{ces96}.
A striking result from these observations is the presence of continuum emission 
in PDRs far from exciting sources \citep{abe02}. This reveals the presence
of VSGs transiently heated by single UV photons.
The continuum relative to the bands is highly variable, showing that VSGs are a dust
component that differs in nature from PAHs \citep{com08}.
Further analysis of the data using principal
component analysis  \citep{rap05a} allows us to extract three fundamental spectra, the various
mixtures of which create the observed spectral variations.
Two of the spectra carry only band emission (they are assigned to
PAH cations and neutrals) and the other one has band and continuum emission
(it is assigned to VSGs). 
Interestingly, the drop of the VSG emission at the cloud edge appears
to be correlated with the increase in the PAH emission, strongly suggesting that free
PAHs are produced by destruction of VSGs under UV irradiation.
This idea is fully consistent
with the strong spatial variations observed in the IRAS 12/100\,$\mu$m ratio at
the surface of molecular clouds \citep{bou90, ber92}. 
These studies continued then taking advantage of the IRS spectrometer onboard
the Spitzer telescope  \citep{hou04}. In particular, the SPECPDR  programme
was dedicated to the study of the evolution of the mid-IR emission spectrum,
by performing 5-35\um\ spectro-imagery of mild-excited PDRs \citep{job05}.
\citet{ber07} analyzed these data using blind signal separation (BSS) analysis.
They confirmed the results previously obtained by \citet{rap05a}
and extended them to a larger set of regions and a larger spectral range.
The two spectra assigned to PAHs were shown to contain bands
only whereas the VSG spectrum to carry the continuum, up to 25\,$\mu$m.
In cool PDRs, as the ones considered in this study with a UV field of the order of
a few 100 G$_0$ or less, the continuum emission is dominated
by stochastically heated dust particles and not by larger grains at thermal equilibrium.

\begin{figure}[h!]
\plottwo{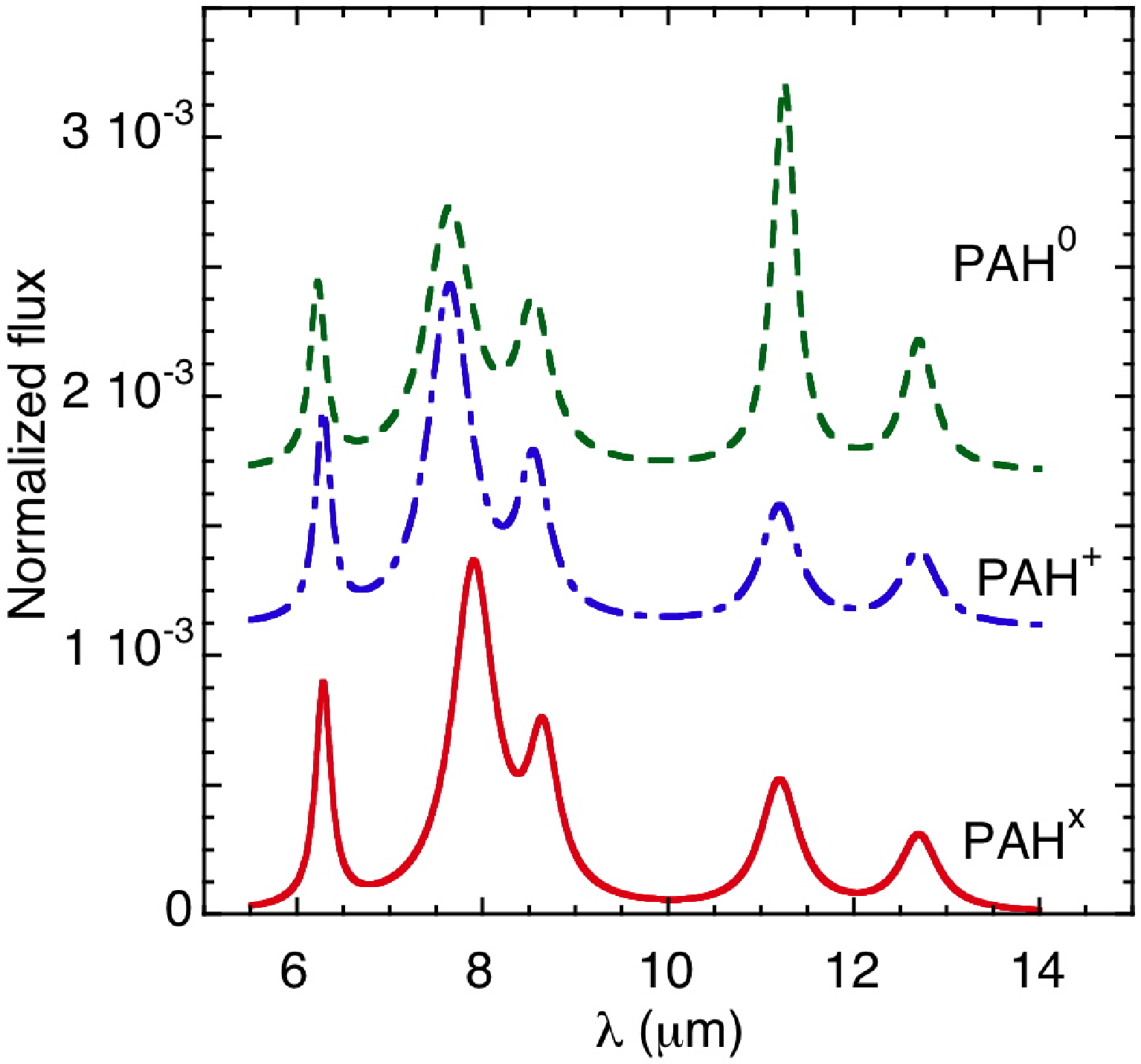}{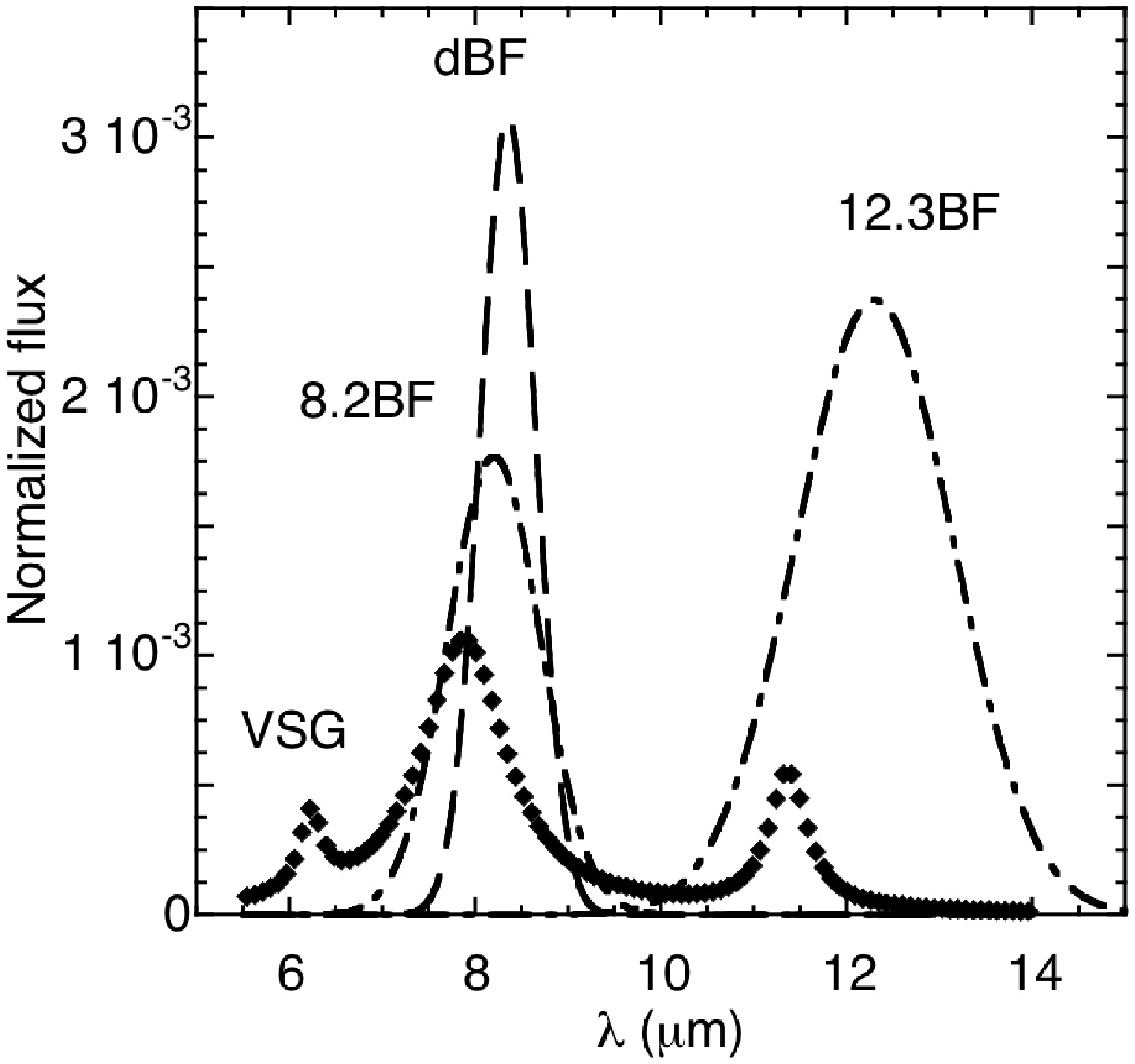}
\caption{The 7 mid-IR template spectra that are used to fit the observed 
emission in mild UV-excited PDRs \citep{rap05a, ber07}, protoplanetary disks \citep{ber09},
planetary nebulae (PNe) and \ion{H}{ii} regions \citep{job08}.
The various PAH mixtures are displayed on the left : PAH$^0$, PAH$^+$ and PAH$^x$
(large ionised species only clearly observed in highly UV-irradiated environments).
The right panel displays the spectrum from VSGs after continuum subtraction as
well as some broad features that are only observed in specific classes
of objects: the broad features at 8.2 and 12.3\um\ in PNe (8.2BF and 12.3BF), and the feature
at 8.3\um\  (dBF) in disks.}
\label{templates}
\end{figure}

The studies on PDRs by \citet{rap05a} and \citet{ber07} led to two
important results. The first one is that free PAHs are efficiently produced by destruction 
of VSGs, most likely under exposure to UV photons. This led \citet{rap05a} to propose
PAH clusters as good candidates for these VSGs. The second important result is the
fact that the mid-IR emission in these regions can be well accounted for by a very
limited number of components: 3 different spectra from 3 different populations, namely
neutral PAH$^0$, cationic PAH$^+$, and VSGs. It is of interest to consider these
spectra to analyze the spectra of other types of objects in particular those in which
UV-excitation conditions are more extreme. This was the motivation to study
the mid-IR emission spectra of planetary nebulae (PNe), \ion{H}{ii} regions
\citep{job08} and protoplanetary disks \citep{ber09}.
In these objects, grains at thermal equilibrium are likely to contribute to the mid-IR continuum.
The description of this continuum would require detailed modelling for each
source, with variations in dust temperature producing variations in its
spectral shape. It cannot, therefore, be reproduced by a simple linear
superposition, with varying weights, of a small number of fixed spectra,
which is the very assumption underlying the applicability of BSS.
A continuum combining two slopes was, therefore, subtracted from the observations
keeping an appropriate level for band wing emission.  
Three PDR-type spectra were built by taking  an average spectrum of each species
from the spectra extracted by \citeauthor{rap05a} and \citeauthor{ber07} in NGC\,7023,
Ced 201, and the $\rho$-Ophiucus filament.
The continuum from VSG spectra was removed since only bands are fitted.
Each average spectrum is then fitted using a combination of Lorentzians
and normalised to their 6-14\um\ integrated intensity.
The resulting template spectra are shown in Figure~\ref{templates}.
All major features have been included except
a plateau at $\lambda \geq 12\,\mu$m for VSGs which was difficult to reproduce due
to contamination by the H$_2$ line at 12.3\um\ and underlying continuum.

The basis consisting of the three PDR components was found to be insufficient
for obtaining a satisfactory fit of the observed spectra in highly-excited objects. 
First, in all classes of objects, PNe,  \ion{H}{ii} regions and disks around A stars,
the presence of another PAH population was evidenced and called PAH$^x$.
The PAH$^x$ spectrum was built empirically to account for the mid-IR emission observed
in PNe and compact \ion{H}{ii} regions \citep{job08} but was inspired by quantum chemistry
calculations \citep{bau08}. 
The main characteristic of this spectrum is that it has a strong feature at 7.9\um\, significantly
redshifted with respect to the 7.6\um\ feature of PAH$^0$ and PAH$^+$. The theoretical calculations
by \cite{bau08} show that such redshift can be explained if one considers
compact PAHs which are large (of the order of 100 or more carbon atoms) and
ionised (cations or anions).
Second, broad emission features (BFs) had to be considered: BFs at 8.2 and 12.3 \um\  that are
inspired from post-AGB spectra \citep[type C objects of][]{pee02}, and a dBF (for disk broad
feature) at 8.3\um\ was extracted from T-Tauri spectra \citep{bou08}.
These features are thought to be associated to material that is rich in aliphatics
and easily destroyed by UV irradiation \citep{slo07}.

\begin{figure}[h!]
\plottwo{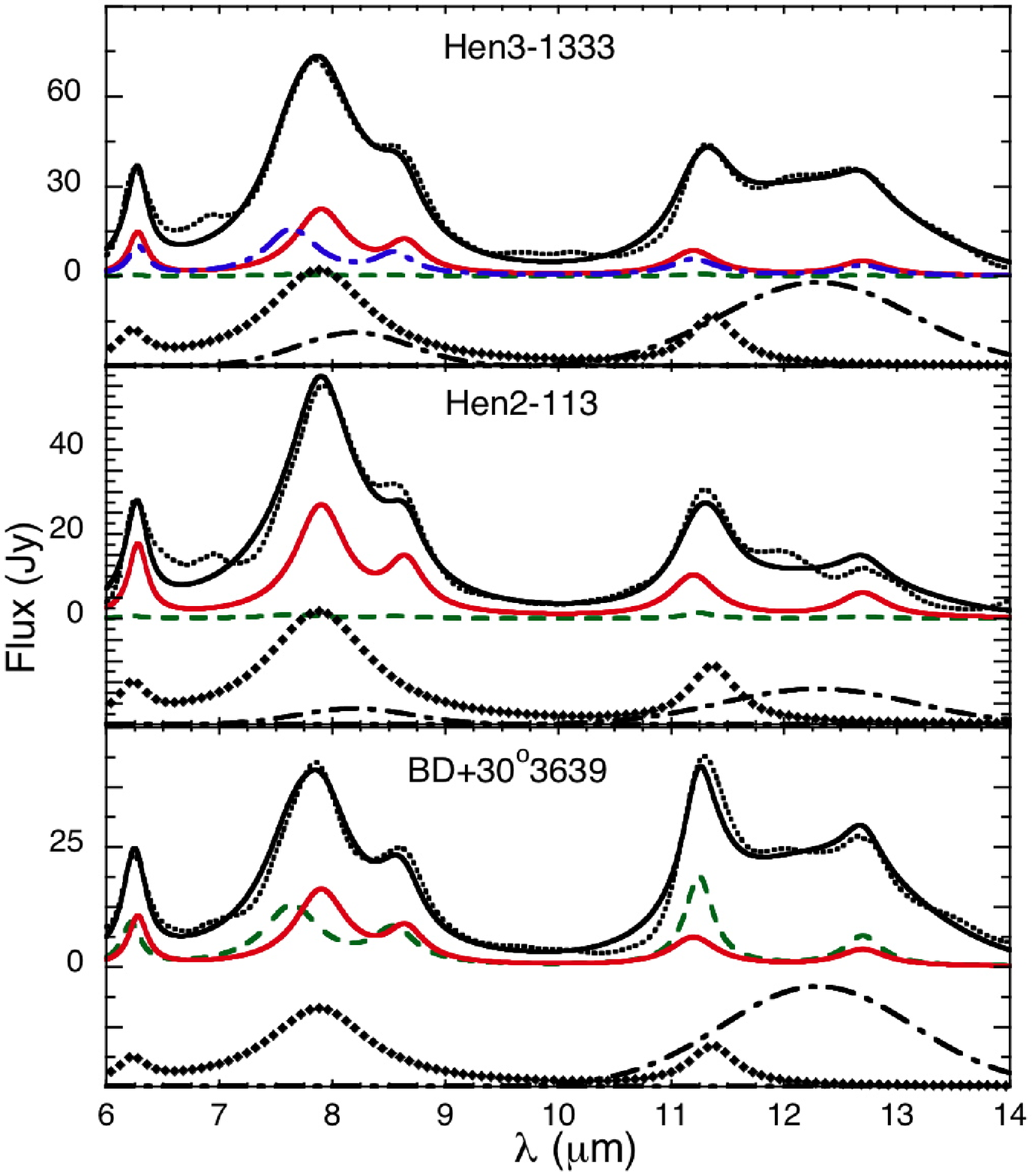}{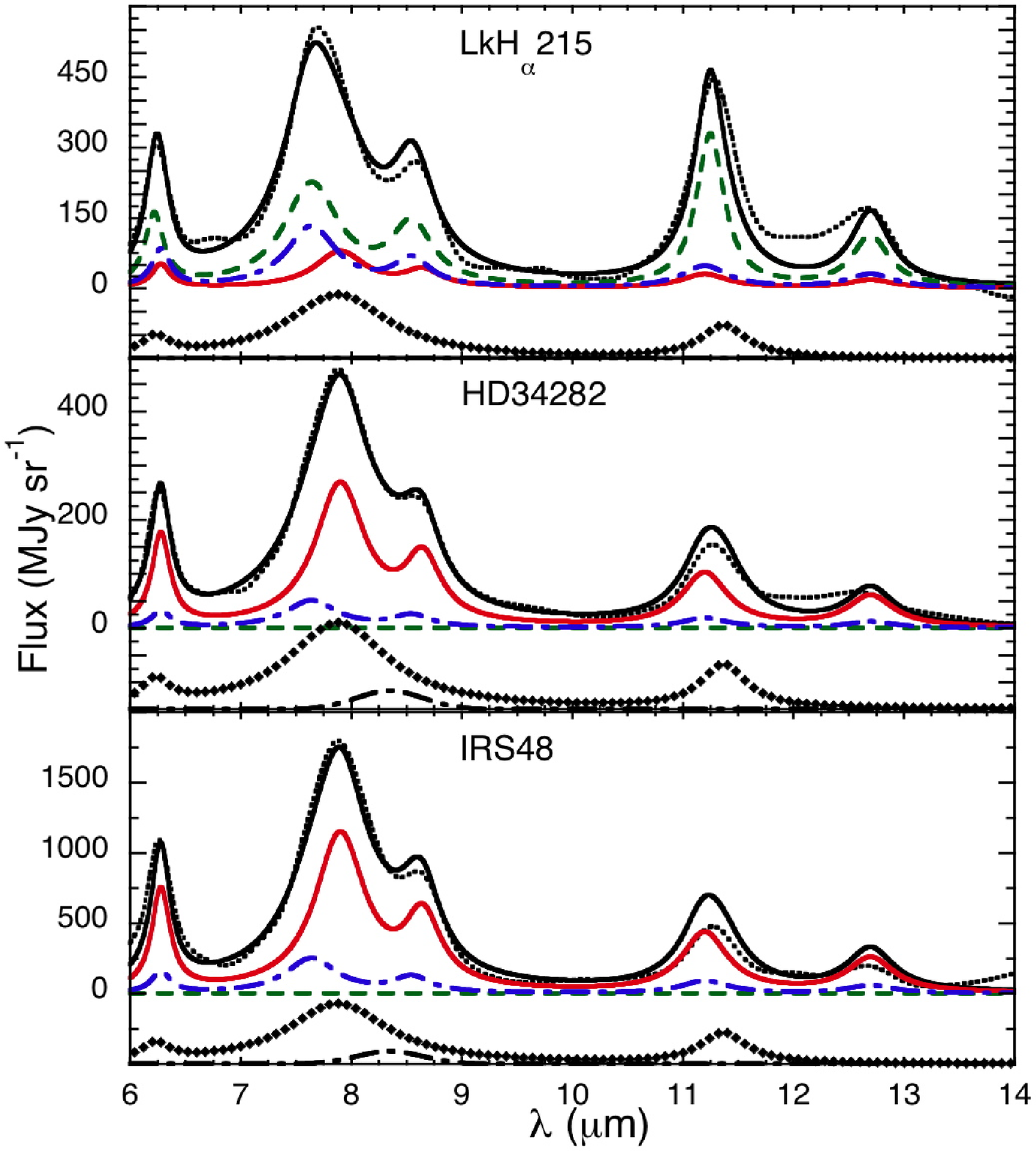}
\caption{Examples of the analysis of the mid-IR spectra of planetary nebulae \citep[left panel; from][]{job08}
and protoplanetary disks \citep[right panel; from][]{ber09}. The spectra after continuum subtraction have been
smoothed at the resolution of $\lambda / \Delta\lambda = 45$ (dashed line) and fitted using the template
spectra displayed in Figure~\ref{templates} (solid line). The PAH components are displayed
in colour lines:  PAH neutrals (green dashed line), cations (blue dash-dot line), and
PAH$^x$ (red continuous line). The broader components have been shifted for clarity: VSGs (diamonds),
8.2, and 12.3\um\  BFs, and dBF (dash-dot line). The ionic gas lines have been removed from the
PNe spectra.}
\label{fits}
\end{figure}

The set of the 6 template spectra displayed in Figure~\ref{templates}: 
PDR-type (3), PAH$^x$,  8.2 and 12.3 \um\ BFs,
were successfully used in \citet{job08} to interpret the mid-IR emission arising from
PNe and \ion{H}{ii} regions both in the Galaxy and Small and Large Magellanic Clouds.
Similarly, the set of the 5 template spectra: PDR-type (3), PAH$^x$, and 8.3 \um\ dBF
were successfully used to analyze the spectra of some protoplanetary disks \citep{ber09}.
The obtained results are in line with the photodestruction of VSGs leading to free PAHs
and the survival of only the largest ionised PAHs, the PAH$^x$ in highly irradiated
environments.  The BFs in PNe are only observed in young sources, which is in line
with efficient destruction of their carriers  by UV irradiation.
 
One of the implications of this work is that the mid-IR emission can be used
to trace the local UV irradiation conditions. The observation of the  PAH$^x$
component reveals strong irradiation conditions.
In protoplanetary disks, \cite{ber09} show that the relative abundance of
the different components: VSG, PAH$^0$ and PAH$^+$, and PAH$^x$
can be used to characterise the central object. The authors also noticed that
PAH$^x$ are not observed around hot Be stars, which strongly suggests that in
these objects the observed emission is dominated by the nebula surrounding
the star and not by the disk. 
This work also provides a rather simple
chemical evolution scheme for mid-IR emitters through the cycle of dust
from evolved stars (PNe), to PDRs associated to molecular
clouds and to the circumstellar environment of young stars including their disks.
The PAH$^0$, PAH$^+$ and VSG components are found in evolved stars and
in PDRs as well. 
VSGs appear as loose aggregates of PAHs. They could be dissociated in
UV-irradiated conditions and reformed in the denser and more shielded
environments of molecular clouds \citep{rap06}.
However, there is no obvious chemical scenario to rebuild PAH
molecules in molecular clouds. This, therefore, puts a constraint on the lifetime
of PAH species considering that at least part of them has to survive the processing
by shocks in the diffuse ISM before being incorporated in molecular clouds
\citep[cf.][ and A. Jones in this volume for a description of the processing
of carbon dust in shocks]{jon96}.
Our study suggests that some PAHs have to survive against these shocks and
constitute the building blocks to regenerate VSGs in molecular clouds.

There is still some effort to be done to explore the conditions in which the
PAH$^x$ component is observed. It was assigned to very large compact
ionised PAHs. As discussed in \citet{job08}, these species are likely
to be present in mild-excited PDRs as well, but their emission is revealed in
highly irradiated environments as due to the combined effect of
destruction of the smaller PAHs (the  PAH$^0$ and PAH$^+$ components)
and higher excitation conditions that bring these very large PAHs to
higher temperatures. In lower irradiation environments, efficient emission in the
6-9\um\ range arises from PAHs of diameter less than 10\AA\  \citep{dra07a}.
Assuming a compact shape for these species, this leads to sizes of less
than 120 carbon atoms \citep[cf. Eq. 2 in][]{omo86}.
Larger PAHs will emit efficiently in this range if they can reach higher internal
temperatures, e.g. contain more than 13.6\,eV of internal energy.
This is possible under high enough UV flux, leading to the
absorption of multiple UV photons; the UV absorption rate is then fast enough
to make it likely that one or more UV photons can be absorbed before the
energy from the previously absorbed ones has been completely reemitted.
PAH$^{x}$ could absorb multiple UV photons when very close to the central
star, which is clearly the case in protoplanetary disks \citep{ber09}.  In the case of PNe
or \ion{H}{ii} regions, the rise of internal temperature could be also due to the
absorption of  extreme UV (EUV) photons ($h\nu > 13.6$\,eV) if PAH$^x$ are
located inside the ionised region or very close to it.
 \citet{com07} measured some PAH emission in the \ion{H}{ii} region facing
the Horsehead nebula, suggesting that these species could indeed be excited by
EUV photons. Further studies at high spatial resolution are, however, necessary
to conclude on the exact location of PAH$^{x}$ in the compact objects studied
in \citet{job08} and \citet{ber09}.

\begin{figure}[h!]
\plotone{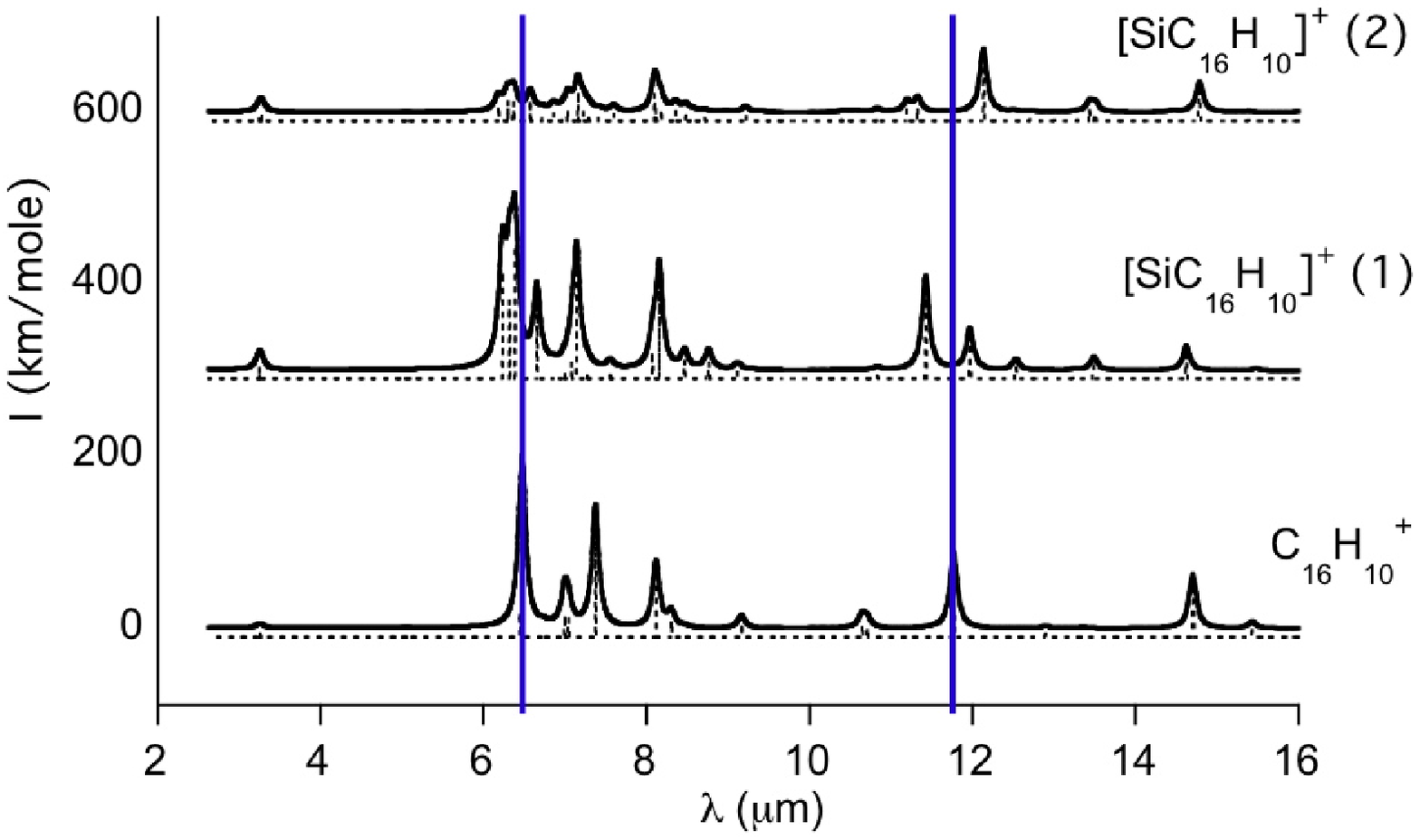}
\plotone{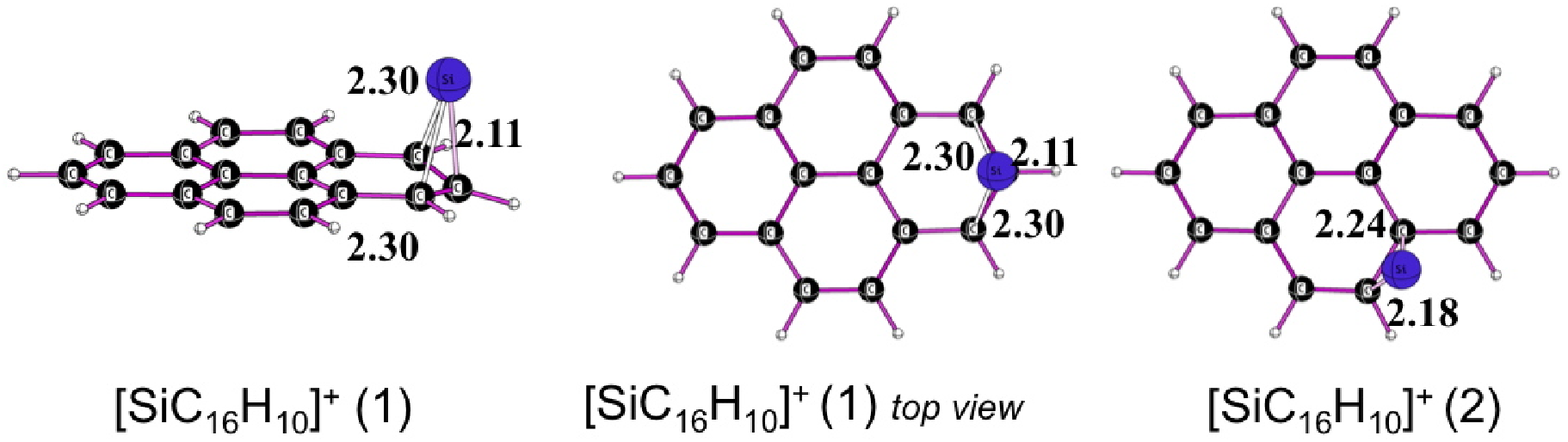}
\caption{Lowest energy structural isomers for the cationic complexes of silicon with
pyrene C$_{16}$H$_{10}$ and corresponding IR spectra, calculated
using a version of the density functional theory \citep[cf.][]{joa09}. The Si-C bond
lengths are given in  \AA.}
\label{SiC16H10}
\end{figure}

\section{The identification of PAHs (and VSGs)}   

\subsection{The spectroscopic view}

\begin{figure}[h!]
\plottwo{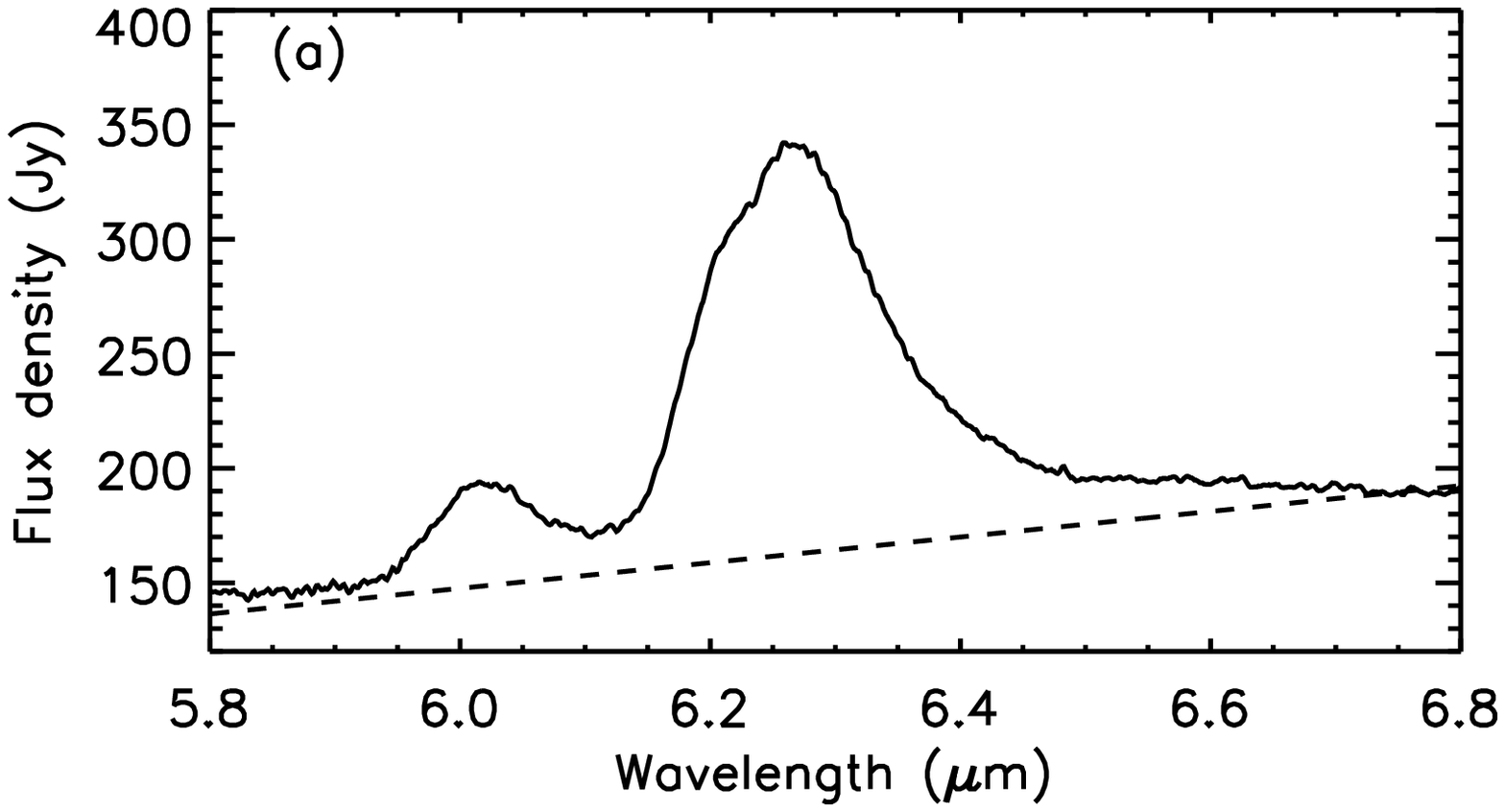}{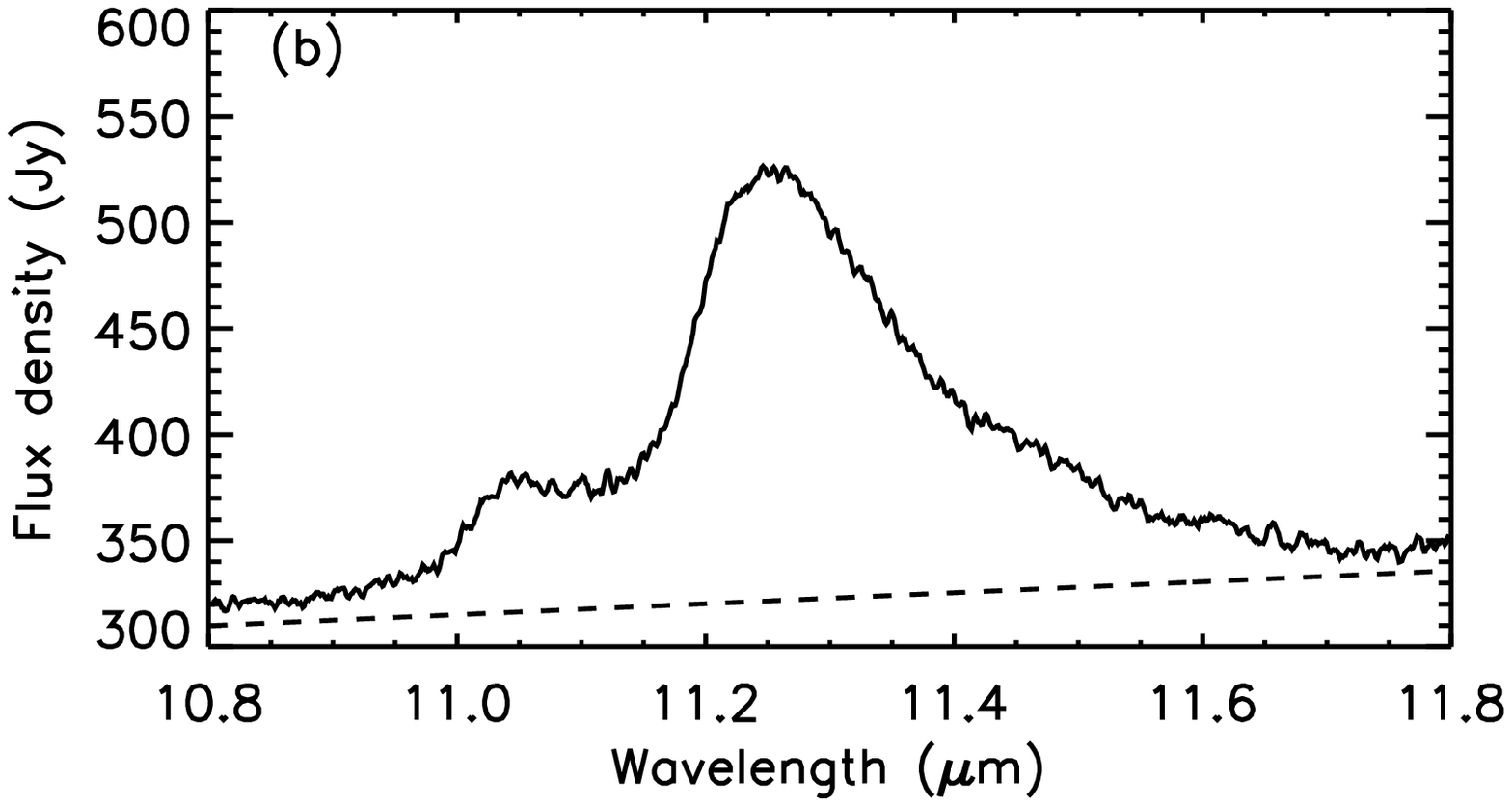}
\caption{The IR spectrum of the Red Rectangle nebula observed with SWS and
retrieved from the ISO archive.
\citet[][see also Figure~\ref{SiC16H10}]{joa09} showed that
[SiPAH]$^+$ $\pi$-complexes have specific mid-IR fingerprints
that can provide an explanation for -(i)- the 6.2 $\mu$m component of the
" 6.2 $\mu$m" AIB and -(ii)- the presence of satellite bands observed on the blue
side of the 6.2 and 11.2 $\mu$m AIBs. From this assignment, it is deduced that
typically 1\% of the cosmic silicon is attached to PAHs.}
\label{RR}
\end{figure}

The identification of interstellar and circumstellar PAHs and VSGs relies on the
match between their spectroscopic fingerprints and bands that are observed in astronomical
spectra. The mid-IR range reflects mainly chemical bonds and, therefore, provides
information on the composition and some trends about the size and structure, but
is not well suited to identify individual species.
As discussed in Section\,\ref{IRobs}, three main populations
were extracted by \citet{rap05a}, \citet{ber07} and \citet{job08}: the PAH$^0$, PAH$^+$ and PAH$^{x}$.
Detailed studies on the positions and shapes of specific bands can also provide
insights into the nature of the emitting species. For instance, \citet{hud05} showed
that the blueshifted position of the 6.2\um\ band cannot be accounted for by PAHs,
neutrals or cations. The authors put forward an alternative explanation involving N atoms
inserted within the C-skeleton. Recently, \citet{joa09} proposed another scenario
in which this band is the signature of [SiPAH]$^+$ $\pi$-complexes in the ISM,
emphasising the importance of the interaction of heavy atoms with PAHs
(cf. Figures~\ref{SiC16H10} and \ref{RR}).

\begin{figure}[hb!]
\plotone{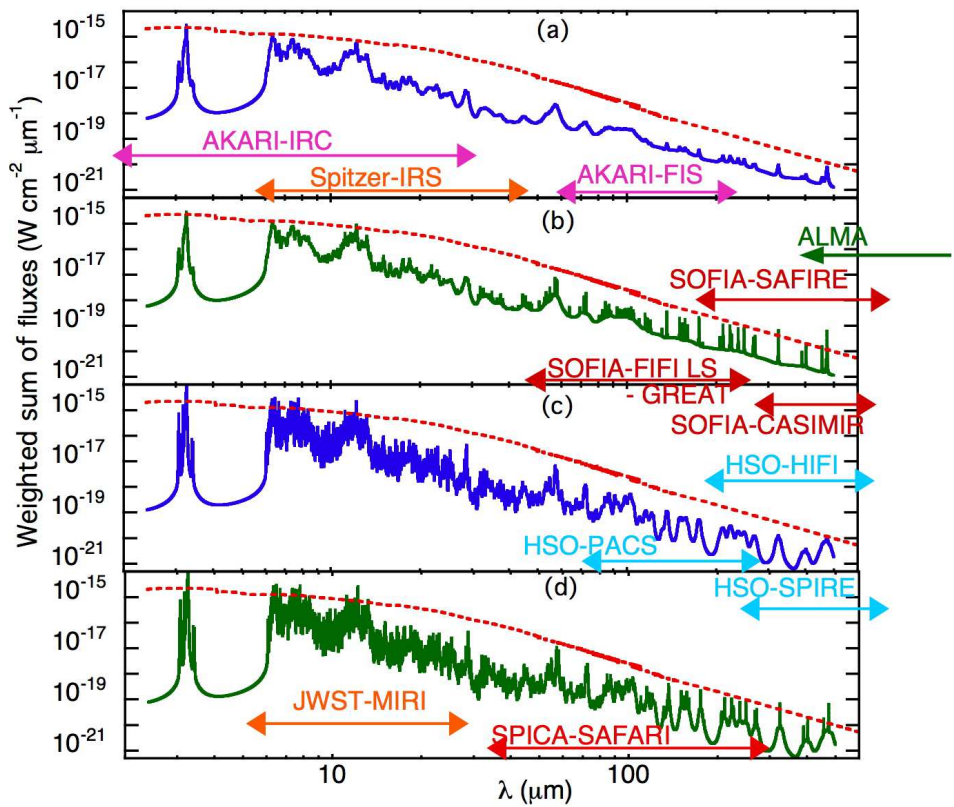}
\caption{Calculated infrared emission spectrum from a sample
of 20 PAHs ranging in size from $C_{10}$H$_8$ to $C_{48}$H$_{20}$,
each in two charge states  for a total of 40 distinct spectra, compared with the
estimated dust continuum in the Red Rectangle nebula.
Different panels correspond to
different assumptions for the band profiles,
$\sigma_{PR}$ representing the width of the P and R rotational branches
and $\sigma_{Q}$ the width of the central Q branch, which is assumed to depend
on the band frequency $\omega$ (see \citet{mul06c} for details).
(a) $\sigma_{PR}= 5.5\,cm^{-1}$, $\sigma_{Q}/\omega=0.7\%$, 
(b) $\sigma_{PR}= 5.5\,cm^{-1}$, $\sigma_{Q}/\omega=0.07\%$, 
(c) $\sigma_{PR}= 0.7\,cm^{-1}$, $\sigma_{Q}/\omega=0.7\%$, 
(d) $\sigma_{PR}= 0.7\,cm^{-1}$, $\sigma_{Q}/\omega=0.07\%$, 
The operation ranges of the relevant astronomical facilities are overlaid
for reference. The HSO will open the possibility to identify PAHs
by their far-IR bands.}
\label{FIR}
\end{figure}

In the UV range, PAHs and VSGs are expected to contribute to both the 220 nm bump
and the far-UV rise.  \citet{cec08} have shown that the two spectral features and
their variations can be accounted for by mixtures of PAHs in different ionisation states.
Whereas the composition of the mixtures is not unique, the trend given by the variation
of the charge state is robust. However, this model only considers free PAHs
due to the lack of spectral data on VSGs whether they are PAH clusters or of another nature.
Diffuse interstellar bands (DIBs) that fall essentially in the visible range \citep{her95}
are a way to identify individual PAH molecules by their low-lying electronic transitions,
which are very characteristic. The fishing for candidates has not been successful so far
\citep[see for instance][]{sal99}. Only \citet{igl08} recently claimed a tentative
identification of ionised naphthalene C$_{10}$H$_8^+$ along one line of sight.
Rotational spectroscopy is another way to identify specific candidates with the limitation that
this applies only to species with a significant value of their dipole moment.
\citet{lovas05} proposed that corannulene, C$_{20}$H$_{10}$, is a good candidate
considering the large value (for a PAH) of 2.07\,D for its dipole
moment. A recent study \citep{pil09}  combines
spectroscopic results obtained in the laboratory (Herbeth, Giesen, in preparation),
numerical simulations of the emission of corannulene in astronomical environments
using a dedicated Monte\textendash Carlo code \citep{mul06c}, and astronomical
observations at the IRAM 30\,m telecope.
An upper limit for the abundance of corannulene in the Red Rectangle
was derived. It amounts to 10$^{-5}$ for the carbon abundance in this species
relative to the total carbon contained in PAHs.

An alternative for the spectroscopic identification of PAHs is a search for their emission
features in the far\textendash IR range, these features being much more specific to the exact
molecular identity than the ones in the mid\textendash IR. The far-IR bands are emitted
during the cooling cascade of UV-excited PAHs, mainly at the end of the cooling
\citep{job02a}.
The emission spectrum of a large sample of PAHs in different charge states
(neutrals and cations) and in specific objects (such as the Red Rectangle
nebula) has been computed by \citet{mul06c}. The
Monte\textendash Carlo code takes as an input molecular parameters
that are available in the theoretical on\textendash line spectral database
of PAHs \citep{mal07b}:  the positions and intensities of all vibrational modes
and the photo\textendash absorption cross\textendash sections up to the
vacuum\textendash UV. The results presented in  \citet{mul06c} concern the
intensities emitted in the different bands.
In Figure~\ref{FIR}, different widths have been assumed to account for the PQR
rotational structures of the bands.
Comparison with the observed spectra however requires a
detailed calculation of the emission profile. 
During the cooling cascade
of the excited PAH, all but the very last one of the transitions are 
not a fundamental (i.~e. $1\rightarrow 0$) transition, but instead happen
between two vibrationally excited states.
The peak position of the IR bands is a function of both the coupling between modes
leading to a cross-anharmonic shift and the quantum numbers in the emitting
vibrational mode leading to an anharmonic shift \citep{oom03}.
The effect of vibrational anharmonicity on band shape for each band requires
then the knowledge of all the molecular parameters, which are not easy to access
by experiment or theory \citep{mul06b}.
A recent study has been performed on naphthalene, C$_{10}$H$_8$ \citep{pir09},
that illustrates  the interplay between high-resolution spectroscopy and numerical
simulations of the photophysics of PAHs. The authors have shown that the
model developed by Mulas et al. is able to predict the 
overall band shape when rotational structure and hot band sequences are
introduced in the modelling. Extension of this work to larger molecules is expected, in
the future, to benefit on the experimental side from the AILES far-IR beamline at the
SOLEIL synchrotron facility.
From an observational point of view, the future will be the search for the far-IR bands
of PAHs with the HSO. The best strategy to detect
these bands is to search for the Q branches associated to these
out\textendash of\textendash plane modes. The PACS
and SPIRE instruments provide the best compromise between
resolving power and sensitivity to evidence these features.  Follow-up
observations at very high resolution with the heterodyne
spectrometer HIFI will allow us to resolve the hot band structure of the Q branches
and may be also some structure in the P and R branches. From these data,
detailed information on the emitting molecule will be retrieved.

\subsection{On the nature of dissociating VSGs}

\begin{figure}[h!]
\plotone{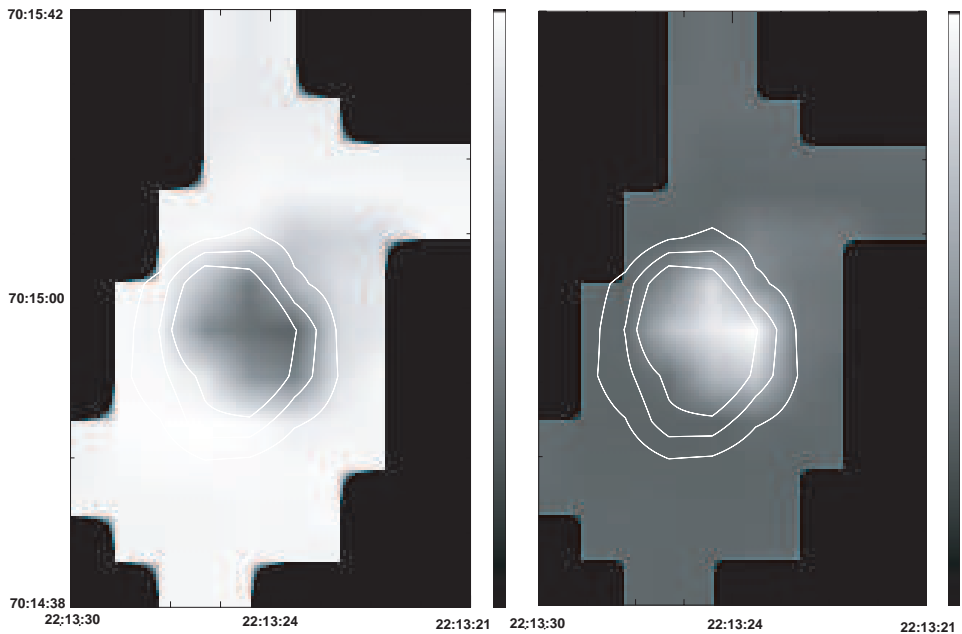}
\caption{Spatial distribution of the emission of the PAH (PAH$^0$, PAH$^+$)
on the left panel and VSG (right panel) components extracted in Ced201 by
\citet{ber07}. Contours represent the integrated mid-IR (5-35\um) emission.
This illustrates the chemical evolution from VSGs to PAHs due
to UV radiation from the star situated in the center of the field of view. }
\label{Ced201}
\end{figure}

As discussed in Section~\ref{IRobs}, free PAHs are produced by destruction of
VSGs. This process is observed in regions where UV photons are present,
including PDRs with mild excitation conditions, in which the chemical frontier
between PAHs and VSGs can be resolved (cf. for instance Figure~\ref{Ced201}).
This implies that VSGs dissociate
quite easily and are, therefore, rather loose aggregates.
\citet{rap05a} proposed that they are clusters of PAHs with a minimum size of
typically 400 carbon atoms per cluster. Taking into account the fact that these
VSGs are the carriers of the continuum up to 25\um\ in these regions
\citep{ber07}, some of the dissociating VSGs could
contain up to a few thousands of carbon atoms.

The formation and destruction mechanisms of 
(C$_{24}$H$_{12}$)$_4$ and (C$_{24}$H$_{12}$)$_{13}$ have been modelled
in detail by \citet{rap06} in the northern PDR of NGC\,7023. 
This study provides a description of the formation mechanism
by molecular dynamics simulations and of the photodissociation by using
a statistical theory (phase space theory). The description of the photophysics of
a specific cluster in the astronomical environment  considered is then provided
using a kinetic Monte\textendash Carlo code that follows \citet{job02a}.
This work led to the conclusion that the studied neutral PAH clusters
are destroyed more rapidly than they are formed in the NGC\,7023 northern PDR.
This is consistent with emitting VSGs being quite larger than the studied
species. There are also several other factors that can reinforce the stability of the clusters
including a larger size for the PAH units \citep{rap05b} and charge effects \citep{rap09}.
There is also the possibility for a  continuous source of production
of VSGs in PDRs due to advection processes bringing unprocessed matter
from the molecular cloud into the UV-exposed layers.
All this leads to different scenarios that have to be explored, putting forward
both fundamental data and astronomical models.

The idea for the presence of PAH clusters has been strengthened  by the observed
spatial correlation between the region where VSGs dissociate and
that where extended red emission (ERE) arises \citep{ber08a}.
This suggests that PAH clusters are ERE carrier precursors.
Furthermore, theoretical studies \citep{rhe07} show
that charged PAH dimers (PAH$_{2}^+$) can fluoresce efficiently 
in the 600 to 1000 nm range. Although this concerns only some specific dimers that
are closed-shells, \citet{ber08a} have shown that this is still consistent
with the observations, considering that ERE carriers contribute typically to only
a few \% of the mid-IR emission. They also discussed the fact the charged PAH
dimers are quite likely to be present  in PDRs since they constitute the last
step in the dissociation cascade of PAH clusters and they can be reformed efficiently
by collision between neutral and cationic PAHs. Furthermore their stability is increased
relative to their neutral counterpart \citep{rap09}.

Serra et al. (1992) and Chaudret et al. (1991) suggested that interstellar PAHs can coordinate
efficiently to metal atoms and in particular to iron. Modelling work by \citet{mar94} suggests that
VSGs made of PAHs and Fe atoms can be formed and survive in PDRs at least in some rather
shielded conditions (A$_V$$\sim$4-5). This work however suffers from the lack of data
on the fundamental properties (photodissociation, reactivity) of these systems.
In a  recent work, \citet{sim09} found that the presence of Fe atoms in coronene clusters
increases the stability of the aggregate.
The bonding dissociation energy (BDE) of the coronene dimer (C$_{24}$H$_{12}$)$_2$
is calculated to be $\sim$ 1 eV  \citep{rap05b}  and $\sim$ 1.3 eV for its positively charged
counterpart \citep{rap09}.
This BDE was found to be 1.7 and 2.0 eV for [Fe(C$_2$$_4$H$_1$$_2$)$_2$]$^{+}$
and  [Fe$_2$(C$_2$$_4$H$_1$$_2$)$_2$]$^{+}$, respectively \citep{sim09}.
The presence of Fe atoms in the ionised aggregates is, therefore, expected to further
increase their stability compared to the homogeneous clusters, and consequently
increase their lifetime in PDRs.

\subsection{The depletion of heavy elements} 
\label{depletion}

Several authors suggested that part of the iron and other heavy metals is in the form of atoms
 that attach and desorb from very small carbonaceous grains \citep[][see also contributions
from B. Draine and S. Zhukovska in this volume] {wei99, rod02, whi03, rod05}.
The photodissociation of very small iron-PAH grains was proposed to account
for the increased abundance of atomic iron in \ion{H}{ii} regions \citep{rod02}.
The abundances of Fe and Si in translucent lines of sight were recently discussed
\citep{mil07}. This work shows that more of these elements
are in the dust phase when the non-linear far-UV rise of the extinction curve increases,
giving support to the fact that they are accreted by the smallest grain populations, e.g PAHs and VSGs
which are responsible for this rise  \citep{cec08}.
Abundances of gas-phase elements have been derived in the diffuse ISM by UV absorption measurements
\citep{sof94, fit96, sav96, mil07} and in PDRs by analysis of mid-IR emission in atomic forbidden lines
\citep{fue00, kau06, oka08}.  The general conclusion is that the abundance of Fe in the gas-phase is typically
a factor of 5-10 less than that of Si. This means that the probability of collision of a PAH with a Si atom is about
one order of magnitude larger compared to a collision with an Fe atom. But this does not tell how much of Fe
(or Si) is indeed attached to PAHs or VSGs since this results from the competition between attachment
and dissociation, which requires a proper description of the physical conditions in the studied regions
and some knowledge on the fundamental properties of the involved species.

Recently, the presence of Si-PAH$^+$ $\pi$-complexes in PDRs was proposed on the basis
of spectroscopic fingerprints \citep{joa09}. The formation of these complexes is expected to occur
without any barrier by simple radiative association releasing 3.0 eV energy.
A different question is that of the substitution of C atoms by Si atoms in the PAH skeleton
which has still to be adressed.
The Si-PAH$^+$ $\pi$-complexes would involve about
1 \% of the cosmic abundance of Si in the Red Rectangle nebula.
Despite detailed theoretical study, no specific mid-IR spectroscopic fingerprints were
found that could be used to reveal the presence of Fe-PAH$^+$ species in the ISM \citep{sim07}.
The main effect of iron was found to be a charge effect, leading to changes of the band intensity
ratios. Still Fe-PAH$^+$ are expected to live longer in the ISM conditions, considering that
the associated BDE is typically 2.6 eV \citep{sim07} to be compared to 1.5\,eV for Si-PAH$^+$
\citep{joa09}. In addition, we have to consider that Fe atoms have a lot of electrons
available for chemical bonding and the formation of larger [Fe$_x$PAH$_y$] clusters
with $x$ and $y > 1$ is expected. This is not the case for Si.
Experimental studies show a fast formation in the gas-phase
of the  [Fe(C$_{24}$H$_{12}$)$_2$]$^+$ complex and no formation of the
[Si(C$_{24}$H$_{12}$)$_2$]$^+$ complex \citep{poz97}. Therefore  [Fe$_x$PAH$_y$] clusters
are good candidates to consider for interstellar VSGs and deserve some studies to characterize
their properties including their IR spectroscopy.

A crude estimate of the composition of Fe-PAH complexes can be made as follows.
In \ion{H}{ii} regions, both VSGs and PAHs are expected to be destroyed and, therefore,
Fe atoms attached to these species entirely released in the gas phase. 
(cf. the results of \citeauthor{cla95} 1995 on the Rosette Nebula and $\lambda$
Orionis \ion{H}{ii}  regions and the recent work of \citeauthor{pov07} 2007
on the massive star formation region M17).
Currently available values for gas-phase Fe abundance in \ion{H}{ii} regions are
quite inaccurate: 2-30\% \citep{rod02} and $<22\%$ \citep{oka08} of the solar abundance.
If we take a value of 5\% for Fe attached to PAHs and VSGs, which is in fact consistent
with the value given earlier by \citet{klo95}, we obtain an average proportion of 2 Fe
atoms for one PAH in Fe-PAH complexes
\citep[assuming an abundance of $7~10^{-7}$ for PAHs;][]{joa09}.

\section{The laboratory approach}
The analysis of space observations shows that further studies are mandatory to characterize
the properties of PAHs and VSGs and related species including [Si-PAH]$^+$ and
[Fe-PAH]$^+$ complexes in conditions which mimic those found in interstellar space.
The PIRENEA set-up in Toulouse has been developed for this purpose \citep{job02b}.
It combines the electromagnetic trapping and the mass spectrometry performances
of an ion cyclotron resonance (ICR) cell with cryogenic cooling.
The achieved ultra-high vacuum ($\sim 10^{-10}$\,mbar) and the cold temperatures
(30\,K on the trap) allow us to approach the physical conditions of isolation
encountered in interstellar space.
The set-up is equipped with (i) a gas injection interface for reactivity studies and (ii)
photon sources including tunable lasers and a Xe arc lamp that provides continuous
radiation from 200\,nm to the near-IR. The ions are produced by laser ablation-ionisation
from a solid target and are trapped in the ICR cell under the conjugated action of  a
magnetic field provided by a 5\,T superconductor magnet and electrical potentials
applied on the trapping plates.
The ion signal that is used for mass analysis is generated by exciting coherently 
the ion cloud with an alternating electric field that contains the cyclotron frequencies of the ions.
The ion motion induces an image charge signal on the detection electrodes
and this transient signal is recorded and analyzed by Fourier Transform (FT) to retrieve the
cyclotron frequencies of the ions and therefore allowing to derive
their mass-over-charge ratio, m/z \citep{mar92}.
Another advantage of the excitation technique is the ability of isolating, in the ICR cell,
only the species of interest at a given m/z ratio by selective ejection of the other trapped species
\citep{gua96}. Very often, the laser ablation technique generates a mixture of
species including isotopomers and fragments.
It is, therefore, mandatory to isolate the ions of interest for a specific study of their physical
and chemical properties. The ions of interest are ions at a given m/z ratio that
can include different isomeric forms (species with the same chemical formula
but different structures).
The experimental procedure consists, therefore, in (i) the production and isolation of the ions
of interest, (ii) cooling down of the ions either radiatively in the environment of the cold
walls or by collisions with a buffer gas, and (iii) photon irradiation or gas injection
depending on the process that is studied.

 \begin{figure}[h!]
\plotone{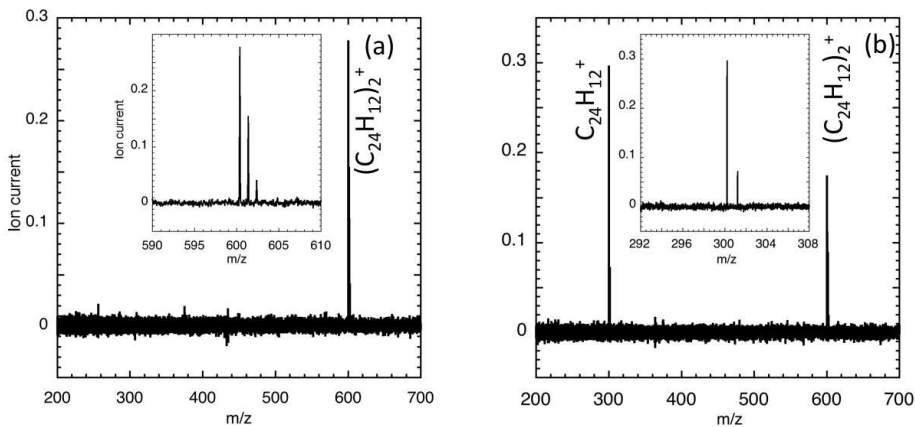}
\caption{Production and isolation of the coronene dimer [(C$_{24}$H$_{12}$)$_2$]$^+$
(m/z=600 and  601, 602  for the $^{13}$C isotopomers) in PIRENEA (a) and dissociation
under irradiation with the Xe arc lamp and the $\lambda_c$=475 nm filter (b).
The dissociation results in the production of the monomer C$_{24}$H$_{12}^+$
(m/z=300 and 301 for the $^{13}$C isotopomer).}
\label{dimer}
\end{figure}

We report here some results that have been obtained on the photodissociation
of laboratory analogues of VSGs  by irradiation with the Xe arc lamp. 
The general question is whether the considered species can easily release PAHs units
under photon irradiation in the gas-phase.
The wavelength range of the Xe lamp  can be adjusted by using longpass colour filters
whose cut-off wavelength $\lambda_c$ is defined at 50$\%$ of transmission. In the experiments discussed
below, we used the $\lambda_c$=475 nm filter.
The dissociation products are recorded by mass spectrometry and kinetics curves can be
obtained by reporting the relative abundances of the parent ion and its photofragments
as a function of the irradiation time which varies from hundreds of milliseconds to several
seconds. 

The first species that are considered are PAH clusters. 
\citet{bre05} reported the formation of coronene clusters containing up to 13 units
in a gas aggregation source. The clusters were ionised before detection in a time-of-flight
mass spectrometer. The limitation of this experiment is that the formed clusters
cannot be handled for further studies. Still, there was the possibility to irradiate the
flying clusters with a focused laser light at $h\nu=$4\,eV, which induces both ionisation
and heating of the clusters. Interestingly, the authors suggest that this process induces
a growth mechanism for the aromatic network. The process was reconsidered by
\citet{sch06} who concluded that the destruction of the cluster should proceed through
evaporation of van-der-Waals bond coronene units and not by intracluster chemical
reactions. More experimental data is, therefore, needed to study the dissociation process
under better controlled conditions.
The PIRENEA set-up is not equipped with an aggregation source.
The dimer [(C$_{24}$H$_{12}$)$_2$]$^+$ was formed in situ in the ICR cell by reactivity of
trapped C$_{24}$H$_{12}$$^+$ with a neutral vapour of coronene.
The formed dimer was isolated (Figure~\ref{dimer} - left panel) and 
irradiated using the Xe arc lamp and the $\lambda_c$=475\,nm filter
(Figure~\ref{dimer} - right panel), showing that the dissociation occurs through
evaporation of a coronene unit.

 \begin{figure}[h!]
\plotone{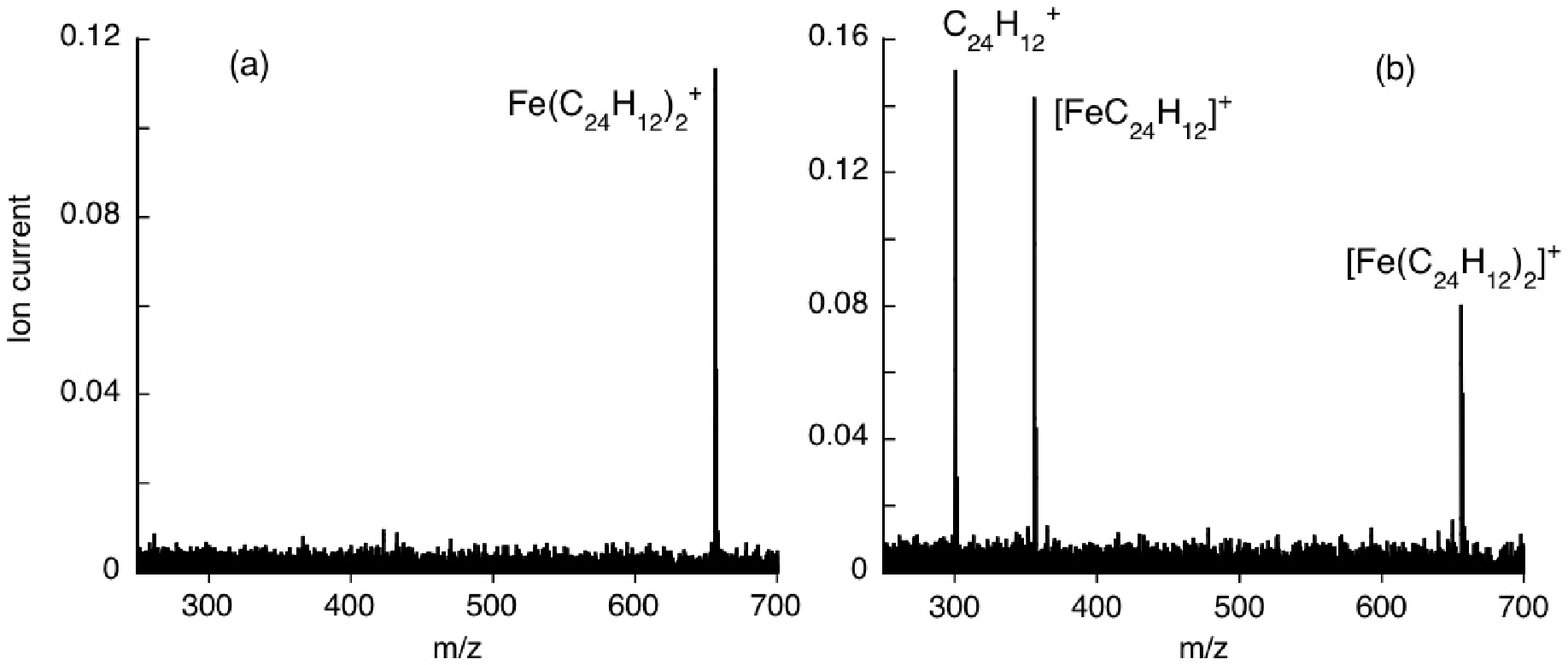}
\plotone{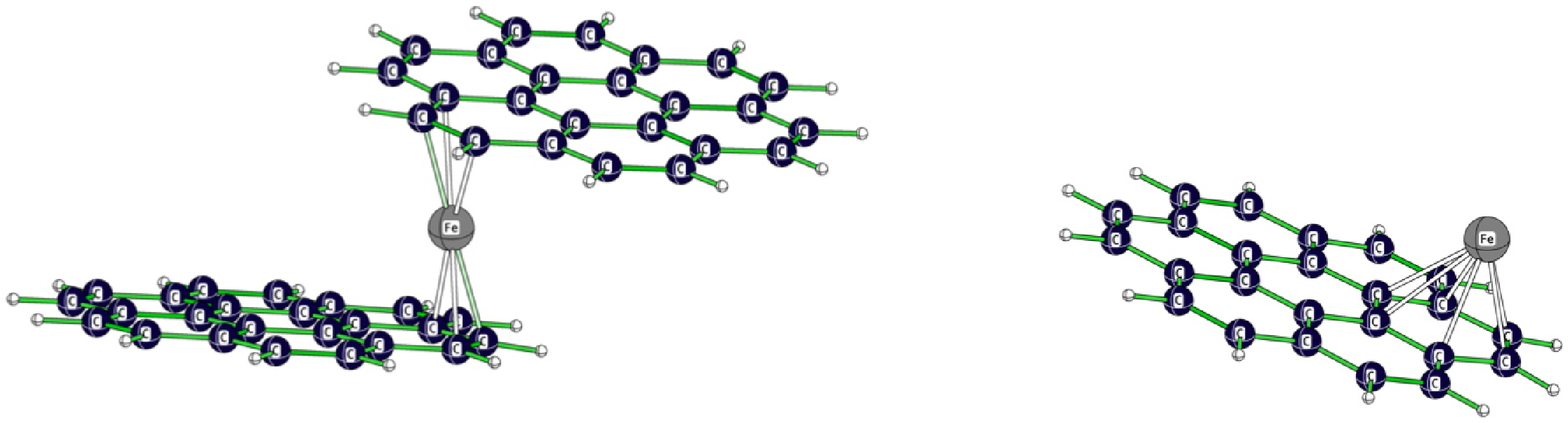}
\caption{Production and isolation of the   [Fe(C$_{24}$H$_{12}$)$_2$]$^+$ complex
(m/z=656) in PIRENEA (a) and dissociation under irradiation with the Xe arc lamp
and the $\lambda_c$=475 nm filter (b). The dissociation results in the production of the
 [Fe(C$_{24}$H$_{12}$)]$^+$ complex (m/z=356) that produces C$_{24}$H$_{12}$$^+$
 by dissociation (right panel). The derived photodissociation pathway is given
 in Figure~\ref{scheme}. The lowest energy structural isomers for the [Fe(C$_{24}$H$_{12}$)$_2$]$^+$
and  [Fe(C$_{24}$H$_{12}$)]$^+$ complexes were calculated
using a version of the density functional theory \citep[cf.][]{sim09}.}
\label{FeC24H12}
\end{figure}

\begin{figure}[h!]
\plotone{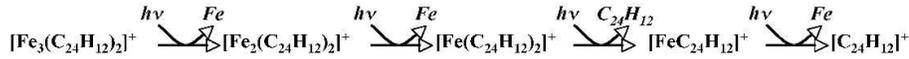}
\caption{The photodissociation sequence for
 the  [Fe$_x$(C$_2$$_4$H$_1$$_2$)$_2$]$^+$ (x=1-3) complexes as observed in the
 PIRENEA set-up.  From \citet{sim09}.}
\label{scheme}
\end{figure}

The other species that are considered are clusters made of PAHs and Fe atoms.
Formation of  [FePAH]$^+$ and [FePAH$_2$]$^+$ by radiative association in an ICR cell
was performed by \citet{dun94} and \citet{poz97}.
A comparable technique could be used in PIRENEA, as for coronene dimers.
Instead, an ablation-ionisation technique was used
with a solid target made of a mixture of triirondodecacarbonyl Fe$_3$(CO)$_1$$_2$
and coronene C$_2$$_4$H$_1$$_2$ deposited onto a thin layer of nanoparticles
of silica \citep{sim09}.
[FeC$_{24}$H$_{12}$]$^+$ and [Fe$_x$(C$_{24}$H$_{12}$)$_2$]$^+$  (x=1-3)
 complexes were formed.
 We discussed in Section\,\ref{depletion} that VSGs could contain an average proportion of 2 Fe
atoms for one PAH. Therefore the species, which have been produced in PIRENEA
can be considered of astrophysical interest. 
 Their photodissociation was recorded under irradiation of the Xe lamp with the
 $\lambda_c$=475\,nm filter. As an example, Figure~\ref{FeC24H12} shows
 the isolation (left panel)  and photodissociation (right panel) of  [Fe(C$_{24}$H$_{12}$)$_2$]$^+$,
 which fits with the general photodissociation pathway reported in Figure~\ref{scheme}.
 [Fe$_x$(C$_{24}$H$_{12}$)$_2$]$^+$  (x=1-3) complexes are found to sequentially
 photodissociate by losing iron atoms and coronene units
 under visible irradiation. The carbon skeleton of coronene is preserved with
 C$_{24}$H$_{12}$$^+$ as the smallest photofragment  \citep{sim09}.
The authors show that (i) the stability of the complex increases with the number of iron atoms, 
and that (ii) the presence of iron atoms induces efficient electronic excitations likely to be responsible 
for a significant increase of the heating efficiency in the experiments conducted with the PIRENEA set-up.

All these experimental results fit into a scenario in which [PAH$_{n}$]  and [Fe$_{x}$PAH$_{y}$]
complexes are formed efficiently in molecular clouds as described by \citet{mar94} and \citet{rap06}
and get easily dissociated upon UV-visible irradiation to produce free PAHs (and Fe atoms/ions).
These are only first results. This work will be pursued by using in particular photon sources
that are more relevant to interstellar conditions. In the experiments described above,
we used visible continuous irradiation, which allows us to build internal energy in the irradiated species
by successive absorption of photons. The dissociation occurs when the dissociation threshold
is reached \citep{boi97}. To mimic photodissociation in PDRs, it is more relevant to use the absorption
of far-UV photons and that will be possible soon in PIRENEA.
It is important to mention that for these large species, the timescale for
relaxing the absorbed energy by IR emission can be larger than the timescale for absorption of UV-visible photons
even in mild UV-excited PDRs. This is indeed this process that is expected to govern the photodissociation of large
PAH clusters in PDRs \citep{rap06}.
 
\section{Summary: The future of PAHs}

The AIBs are a major signature of UV-irradiated environments
and, therefore, could be a valuable tool for astronomers to trace
the local physical and chemical properties in these environments.
In the recent years, some progress has been done in the understanding
of the processing in circumstellar and interstellar environments
of the proposed carriers for AIBs: PAHs and related species (VSGs).
These observational results provide guidelines for laboratory studies. For instance,
an important result is the production of free PAH molecules by photodissociation of VSGs,
which provides insights into the nature of these VSGs.
A still open question is the contribution of PAHs to the depletion of heavy atoms,
such as Si and Fe. Although there is some spectroscopic
evidence for the presence of Si-PAH$^+$ complexes in PDRs, the proof for the adsorption of
Fe on PAHs and its inclusion in VSGs is still lacking. This should be further investigated considering
the potential implication it has in element depletion, charge balance as well as chemistry
(possibility of a catalytic chemistry) in astronomical environments.

A key question remains the identification of individual PAHs through spectroscopic
fingerprints. The identification of electronic transitions by comparison to the DIBs is one way.
Another way will be opened in the near-future with the launch of HSO providing the
possibility to explore the far-IR range where the bending modes of the carbon skeleton
are found. Clearly this subject has some exciting years to go.
Important enough is, from space to the laboratory, to be open-minded in order
to consider different kinds of PAH-related species since the question is not solved yet.
There is still a lot to do at the interface between astronomy and physical chemistry.

\acknowledgements 
We would like to thank the technical team of PIRENEA: M. Armengaud, A. Bonnamy,
P. Frabel and L. Nogues. This work is supported by interdisciplinary programmes :
the CNRS / Programme National Physique et Chimie du Milieu Interstellaire
and the European Research Training Network ÒMolecular UniverseÓ (MRTN-CT-
2004-512302) are especially acknowledged.


\end{document}